\documentclass[pdflatex,sn-mathphys-num]{sn-jnl}

\usepackage{graphicx}%
\usepackage{multirow}%
\usepackage{amsmath,amssymb,amsfonts}%
\usepackage{amsthm}%
\usepackage{mathrsfs}%
\usepackage[title]{appendix}%
\usepackage{xcolor}%
\usepackage{textcomp}%
\usepackage{manyfoot}%
\usepackage{booktabs}%
\usepackage{algorithm}%
\usepackage{algorithmicx}%
\usepackage{algpseudocode}%
\usepackage{listings}%

\theoremstyle{thmstyleone}%
\theoremstyle{thmstyletwo}%

\theoremstyle{thmstylethree}%

\begin{document}

\title[Article Title]{Control-free and efficient integrated photonic neural networks via hardware-aware training and pruning}

\author[1]{Tengji Xu}
\equalcont{These authors contributed equally to this work.}\email{tengjixu@link.cuhk.edu.hk}
\author[2]{Weipeng Zhang}
\equalcont{These authors contributed equally to this work.}\email{weipengz@princeton.edu}
\author[2]{Jiawei Zhang}
\equalcont{These authors contributed equally to this work.}\email{jz8939@princeton.edu}
\author[1]{Zeyu Luo}\email{zeyuluo@cuhk.edu.hk}
\author[1]{Qiarong Xiao}\email{qrxiao@link.cuhk.edu.hk}
\author[1]{Benshan Wang}\email{bswang@link.cuhk.edu.hk}
\author[1]{Mingcheng Luo}\email{mcluo318@link.cuhk.edu.hk}
\author[3]{Xingyuan Xu}\email{xingyuanxu@bupt.edu.cn}
\author[4]{Bhavin J. Shastri}\email{shastri@ieee.org}
\author[2]{Paul R. Prucnal}\email{prucnal@ee.princeton.edu}
\author*[1]{Chaoran Huang}\email{crhuang@ee.cuhk.edu.hk}

\affil*[1]{\orgdiv{Department of Electronic Engineering}, \orgname{The Chinese University of Hong Kong}, \orgaddress{\street{Shatin}, \city{Hong Kong SAR}, \country{China}}}

\affil[2]{\orgdiv{Department of Electrical and Computer Engineering}, \orgname{Princeton University}, \orgaddress{ \city{Princeton}, \postcode{08544}, \state{New Jersey}, \country{USA}}}

\affil[3]{\orgdiv{State Key Lab. of Information Photonics and Optical Communications}, \orgname{Beijing University of Posts and Telecommunications}, \orgaddress{\city{Beijing}, \postcode{100876},  \country{China}}}

\affil[4]{\orgdiv{Department of Physics, Engineering Physics and Astronomy}, \orgname{Queen’s University}, \orgaddress{\city{Kingston}, \postcode{K7L 3N6}, \state{Ontario}, \country{Canada}}}


\abstract{Integrated photonic neural networks (PNNs) are at the forefront of AI computing, leveraging on light's unique properties, such as large bandwidth, low latency, and potentially low power consumption. Nevertheless, the integrated optical components within PNNs are inherently sensitive to external disturbances and thermal interference, which can detrimentally affect computing accuracy and reliability. Current solutions often use complicated control methods, resulting in high hardware complexity impractical for large-scale PNNs. In response, we propose a novel hardware-aware training and pruning approach. The core idea is to train the parameters of a physical neural network towards its noise-robust and energy-efficient region. This innovation enables control-free and energy-efficient photonic computing. Our method is validated across diverse integrated PNN architectures. Through experimental validation, our approach significantly enhances the computing precision of MRR-based PNN, achieving a notable 4-bit improvement without the need for complex device control mechanisms or energy-intensive temperature stabilization circuits. Specifically, it improves the accuracy of experimental handwritten digit classification from 67.0\% to 95.0\%, nearing theoretical limits and achieved without a thermoelectric controller. Additionally, this approach reduces the energy by tenfold. We further extend the validation to various architectures, such as PCM-based PNN, demonstrating the broad applicability of our approach across different platforms. This advancement represents a significant step towards the practical, energy-efficient, and noise-resilient implementation of large-scale integrated PNNs.}




\maketitle

\section{Introduction}\label{sec1}

Neural networks (NNs) are powerful tools with diverse applications, including image recognition, natural language processing, and autonomous driving \cite{lecun2015deep}. However, the ever-increasing task complexity widens the gap between the computational requirements and the processing power of traditional electronic processors. Photonic neural networks (PNNs) emerge as a promising frontier for AI computing by leveraging the unique properties of light, such as large bandwidth, low latency, and potentially low power consumption. PNNs show the potential of bridging the computing gap posed by traditional electronic processors~\cite{shen2017deep,lin2018all,hamerly2019large,huang2021silicon,xu202111,xu2021optical,feldmann2021parallel,zhu2022space,fu2023photonic,ashtiani2022chip,feldmann2019all,wang2023image,zhou2022photonic,filipovich2022silicon,shastri2021photonics,zhou2021large,chen2023all,miscuglio2020photonic,meng2023electrical,peserico2023design,youngblood2023integrated}. They have found applications in diverse computing and signal processing systems ~\cite{huang2021silicon,wang2022multi,zhang2023broadband,zhang2024system}.

However, the accuracy of PNNs can be easily affected by various mechanisms, such as ambient disturbances, particularly those induced by thermal fluctuations, thermal crosstalk from microheaters, limited resolvable weight levels, and device degradation, particularly in phase change materials (PCMs)~\cite{li2024performance}. The challenge is further intensified in large-scale PNNs~\cite{tait2016multi,tait2018feedback,huang2020demonstration, zhang2022silicon,cheng2023self}. For example, we experimentally observe in this work that a small resonance drift caused by thermal fluctuations in a microring (MRR)-based PNN by only tens of pm can cause an accuracy degradation from 99.0\% to only 67.0\% in a small two-layer convolutional neural network (CNN) performing Modified National Institute of Standards and Technology (MNIST) classification task~\cite{lecun1998gradient}. As the size of the PNN further increases, the degradation in accuracy due to resonance drift becomes even more pronounced––the accuracy of the PNN drops from the theoretical value of 83.6\% to 9.15\% when performing Canadian Institute for Advanced Research, 10 classes (CIFAR-10) classification task~\cite{krizhevsky2009learning}. For another example, PCM-based in-memory PNNs encounter challenges related to limited weight resolutions and reproducibility caused by factors including material degradation, stochasticity in the crystallization process, and fluctuations in experimental switching conditions~\cite{li2024performance}. Consequently, the accuracy of a PNN with $\mathrm{Sb}_2 \mathrm{Se}_3$-based phase shifter may reduce from 99.0\% to 63.2\% when the crystallization temperatures vary by 5 K (corresponding to 8 distinguishable weight levels).
These detrimental effects can be addressed by using external electronic control circuits to provide dynamic stabilizations~\cite{li2024performance,huang2020demonstration,zhang2022silicon, cheng2023self,padmaraju2013wavelength,zhu2019self,zhu2014fast,pintus2019pwm,zhang2014towards}. However, such conventional methods incur additional hardware costs and increase system complexity, making it impractical to build large-scale PNNs comprising hundreds or even thousands of components. 

In this work, we introduce a general approach to address all these challenges in the  PNN by introducing a hardware-algorithm codesign approach. This approach enhances the robustness of PNNs while simultaneously reducing power consumption. These improvements are achieved without needing cumbersome control algorithms or introducing additional hardware complexity. Instead, we attain these benefits by intentionally training the PNN towards its noise-robust and energy-efficient region. Our approach shares similarities with "neural network (NN) pruning", a technique commonly used to optimize the energy and memory efficiency in software-based NNs. Pruning in software-based NNs is to remove parameters from the NN model by setting weights to zero. In contrast, our approach focuses on encouraging weights to move to the variance-insensitive region of the device's transfer function, thus improving the robustness of PNNs. Importantly, the variance-insensitive region coincides with the region requiring minimal tuning power for weight assignment. Therefore, our approach also leads to a substantial power consumption reduction. While pruning concepts have been independently proposed in MZI networks, their primary focus is on reducing power consumption without effectively addressing the issue of noise~\cite{sarantoglou2022bayesian,banerjee2023pruning,yu2023heavy,gu2021efficient}.

Our approach solves the problem that PNN requires highly complicated control hardware to achieve and maintain its computing accuracy. We demonstrate that our approach can achieve 4-bit weight precision improvement for a  MRR-based PNN without using thermoelectric controller (TEC) or other dynamic weight control methods. Based on this, the experimental classification accuracy for a two-layer CNN conducting MNIST dataset classification reaches 95.0\%, comparable to the theoretical value. In contrast, the accuracy without using our approach drops to only 67.0\%, due to the resonance drift caused by ambient fluctuations. Moreover, this enhancement occurs alongside 10-fold tuning power consumption reduction. As the NN size expands, the improvement in classification accuracy and reduction in power consumption becomes more pronounced. Using it to tackle a more intricate classification task like CIFAR-10, our approach demonstrates its capability by elevating the accuracy from 9.15\% to 80.6\% with 160 times power reduction. 

We further extend the validation to various architectures, including MRRs with single-end detection~\cite{feldmann2019all,zhang2024compact} and crossbar arrays~\cite{ohno2022si}. Furthermore, our approach effectively solves the problem of limited accuracy in PCM-based PNNs~\cite{feldmann2021parallel} caused by the limited weight resolutions and reproducibility of PCMs. We have demonstrated the improvement of the classification accuracy in MNIST dataset classification from 63.2\% to 98.5\% without re-training the neural network, even in the presence of various PCM degradations. 

Our experiments and simulations show our approach is broadly applicable to influential PNN systems and effectively addresses accuracy reduction challenges caused by various issues, including ambient disturbances, thermal crosstalk, limited resolvable weight levels, and device degradation. Our work represents a pathway towards practical, energy-efficient, and noise-resilient large-scale PNN implementations. 



\section{Operation principle}\label{sec2}

PNNs can be realized by high-speed optical devices that are densely interconnected by parallel waveguides. Synaptic weights in convolutional and fully connected layers can be mathematically transformed into matrix multiplication operations. The matrix can be achieved through either incoherent summations of light intensity or coherent summations resulting from light interference~\cite {shastri2021photonics}. Widely used incoherent approaches include micro-ring resonator (MRR) weight banks ~\cite{tait2016microring,ferreira2022design} and crossbar arrays \cite{feldmann2021parallel}. To apply weights, partial signal transmission is tuned using a phase shifter and added by either a single-end detector or a balanced detector. Compared to coherent approaches such as Mach-Zehnder interferometer (MZI) networks, incoherent PNNs offer advantages in power-efficient tuning, straightforward weight assignment, and high computing density due to the small device footprint~\cite{tait2016microring,feldmann2021parallel,feldmann2019all}. However, the accuracy of weights may be affected by various mechanisms, including ambient disturbances, thermal crosstalk from microheaters, limited resolvable weight levels, and device degradations. Although these issues can be mitigated by retraining the PNNs or by locking weight states using control circuits, these approaches add significant hardware complexity.

To solve these problems, we propose training the weights to a variance-robust region within a PNN. This can be achieved by incorporating a regularization term into the loss function during training. The regularization term serves as a penalty function, encouraging the relocation of weights from variance-sensitive to noise-robust regions:
\begin{equation}\label{eq: regularization_slope}
  L = f(y, \hat{y})+\alpha||\frac{\partial W}{\partial n}||_2
\end{equation}\label{eq:general-lossFunc}
where $L$ is the defined loss function, $f(\cdot)$ is the cross-entropy loss, $y$ is the NN output, $\hat{y}$ is the target label, $\alpha$ is the regularization coefficient, $\frac{\partial W}{\partial n}$ is the slope of weight to the refractive index change caused by various detrimental mechanisms. Here $||A||_2$ denotes the L2 norm of a matrix $A$. $W$ is the weight matrix of each NN layer. By inducing this regularization term, the weights are effectively pushed towards a noise-robust region during training ~\cite{gu2022squeezelight}. Moreover, for certain filter banks such as MRR weight banks, this noise-robust region aligns with areas requiring minimal tuning power and exhibiting the highest resilience to thermal fluctuations. Consequently, our approach offers the following advantages. Firstly, weights are shifted to a region less susceptible to noise and other detrimental mechanisms, resulting in substantially smaller weight errors. Secondly, it significantly reduces power consumption as it requires minimal tuning power. Additionally, by minimizing total tuning power, thermal crosstalk can be effectively mitigated.

\clearpage
\begin{figure*}[ht]
\centering
\includegraphics[width=1\linewidth]{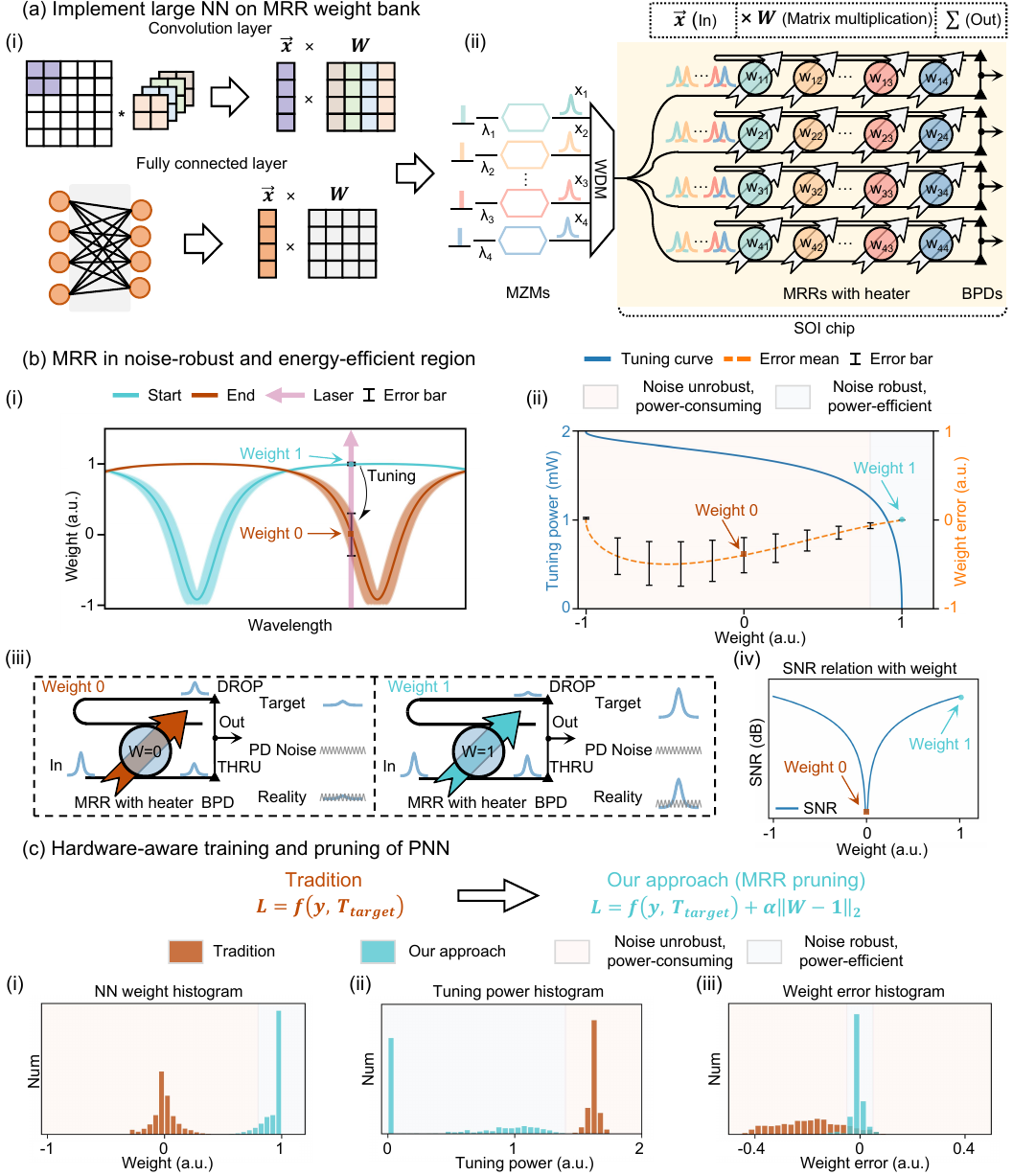}
\caption{Illustration of conducting NN inference on MRR weight bank, MRR operation error performance, and ‘MRR pruning’ optimization method. (a-i) Transform the convolutional layer and fully connected layer into matrix multiplication. (a-ii) Implement matrix multiplication on MRR weight bank. (b-i) Single MRR spectrum change through thermal tuning. (b-ii) Tuning power and weight error change with weight value. (b-iii) Schematic of MRR signal transmission at weight 0 and weight 1. (b-iv) Signal-to-noise ratio change with the weight value. (c-i) Weight value distribution contrast histogram. (c-ii) Tuning power distribution contrast histogram. (c-iii) Weight error distribution contrast histogram. NN: Neural network. MRR: Micro-ring resonator. SOI: Silicon on insulator. BPD: Balanced photodetector. THRU: Through. SNR: Signal-to-noise ratio. PNN: Photonic neural network.}
\label{fig:illustration figure}
\end{figure*}

\clearpage

In the subsequent sections (section~\ref{sec: MRR} and~\ref{section:PCM}) of this paper, we will showcase the application of our approach to different integrated PNN systems. Specifically, in the upcoming section, we provide a comprehensive experimental investigation of MRR-based PNNs, as MRRs are particularly prone to resonance drift. In section \ref{section:PCM}, we further extend the validation to PCM-based PNNs, demonstrating our approach's ability to address device imperfection and degradation-induced accuracy reduction.

\section{MRR-based PNNs} \label{sec: MRR}

\subsection{MRR weight banks with balanced photodetection}

In a MRR-based PNNs, weight synapses can be realized by MRR weight banks, as shown in Fig.\ref{fig:illustration figure}(a-i). The input matrix is represented by an array of analog signals, each modulated onto a laser with a distinct wavelength. The multiplied matrix is realized by an array of MRR weight banks, shown in Fig.\ref{fig:illustration figure}(a-ii). MRRs on each weight bank have a slight radius offset, leading to a distinct resonance wavelength that aligned with the input laser's wavelength. As shown in Fig.\ref{fig:illustration figure}(b-i) and (b-ii), by tuning MRR's resonance wavelength, one can precisely control the portion of light divided into the Drop and Through ports of the MRR, thereby realizing tunable weights. Thermal tuning is the most commonly used mechanism for controlling MRRs due to its easy implementation on Silicon Photonics (SiP) chips. The weighted input signals are subsequently detected by a balanced photodetector (BPD). It subtracts signals from the Through port and Drop port. This results in an effective weight value ranging from -1 to 1, shown in Fig.\ref{fig:illustration figure}(b-iii).  MRR weight banks in other architectures, such as single-end detection and crossbar arrays ~\cite{feldmann2019all,zhang2024compact, ohno2022si} will be discussed in later sections.

We position the input lasers at the edge of each MRR's transmission spectrum, as shown in Fig.\ref{fig:illustration figure}(b-i). In this configuration, the initial weight is 1. At the spectral edge, the weight value of "1" is resilient to the resonance drift caused by the thermal fluctuations. After applying a current to the microheater embedded in the MRR, its resonant wavelength shifts towards a longer wavelength, resulting in a different weight value. Taking weight 0 as an example, weight 0 is realized by tuning the resonance such that the input optical signal is equally divided into the Through and Drop ports. Weight 0 is sensitive to thermal fluctuations, as the signal's wavelength is located at the steep region of the MRR's transfer fluctuations. Fig.\ref{fig:illustration figure}(b-ii), obtained from experiments, plots tuning power and weight error variance as functions of weight values. It is clear that weights near 1 require the least tuning power and are most robust to thermal fluctuations. In contrast, weights near 0 require large tuning power and are sensitive to thermal fluctuations. Furthermore, weight values also lead to different signal-to-noise ratios (SNRs) at the BPD. SNR as a function of various weights is plotted in Fig.\ref{fig:illustration figure}(b-iv), showing weight "1" can provide a much higher SNR as compared to weight "0" considering various noise in BPD (Details are derived in supplement document). Considering all these effects, we call the weights close to 1 as good region (the blue region in Fig.\ref{fig:illustration figure}(b-ii)). It provides advantages of nearly zero tuning power, the highest resilience to thermal fluctuations, and the largest SNRs, simultaneously. By contrast, weights near 0 is bad region (the red region in Fig.\ref{fig:illustration figure}(b-ii)), as it is power-consuming, vulnerable to thermal fluctuations, and suffers the lowest SNR.

In a large-scale PNN, this performance contrast will become even more significant. An important observation through training a NN is, when we use the standard setting in backpropagation to train a NN, most weights are most likely to cluster around 0, the bad region of MRR-based PNN, as shown in Fig.\ref{fig:illustration figure}(c-i). Consequently, a slight resonance drift can detriment the accuracy in a PNN due to error accumulation unless dedicated wavelength-locking is applied to every MRR. However, the associated hardware complexity can be prohibitively high. To solve these problems, we relocate the weights from the bad region to the good region (i.e., weights approaching "1") by incorporating a regularization term into the loss function when training the NNs. In the MRR-based PNN with BPD, the loss function is specified from Eq.~\ref{eq: regularization_slope} to Eq.~\ref{eq: regularization},
\begin{equation}\label{eq: regularization}
  L = f(y, \hat{y})+\alpha||W-1||_2
\end{equation}

To achieve effective weight pruning, we use iterative fine-tuning, a strategy commonly used in software NN, to train the PNN. The strategy involves two steps. The system is first trained using the loss function without the regularization term. Next, the system is fine-tuned using a modified loss function, adding the regularization term. The second step only fine-tunes those weights that are close to zero while keeping other weights frozen~\cite{han2015learning}. This process goes layer by layer. This selective retraining is executed with continuous accuracy monitoring to ensure adherence to performance requirements. After iterative optimization, the majority of weights are moved to 1, the good region. Meanwhile, the computing accuracy stays comparable to the original NN without pruning~\cite{frankle2018lottery}.

Our approach offers notable benefits. First, weights are relocated to the region that is resilient to resonance drift, leading to a much smaller weight error. Second, it leads to a significantly reduced power consumption as weights around 1 require almost zero tuning power. In addition, by reducing the total tuning power, thermal crosstalk could also be effectively mitigated. Finally, the overall SNR of the PNN system will have a substantial improvement. These benefits concurrently improve the accuracy and robustness of MRR-based PNN, without requiring cumbersome control of every MRR in the PNN.

\subsection{Experiment results}
\subsubsection{Realization of control-free and robust PNNs} Our experiments primarily involve two key aspects, the first is to validate the effectiveness of our approach in reducing weight error. Next is to determine how our approach can significantly enhance NN inference classification accuracy. Here we assess the effectiveness of our approach across three distinct NNs of varying sizes and for different tasks, chosen according to a classic NN pruning work~\cite{frankle2018lottery}.

\begin{figure*}[ht]
\centering
\includegraphics[width=1\linewidth]{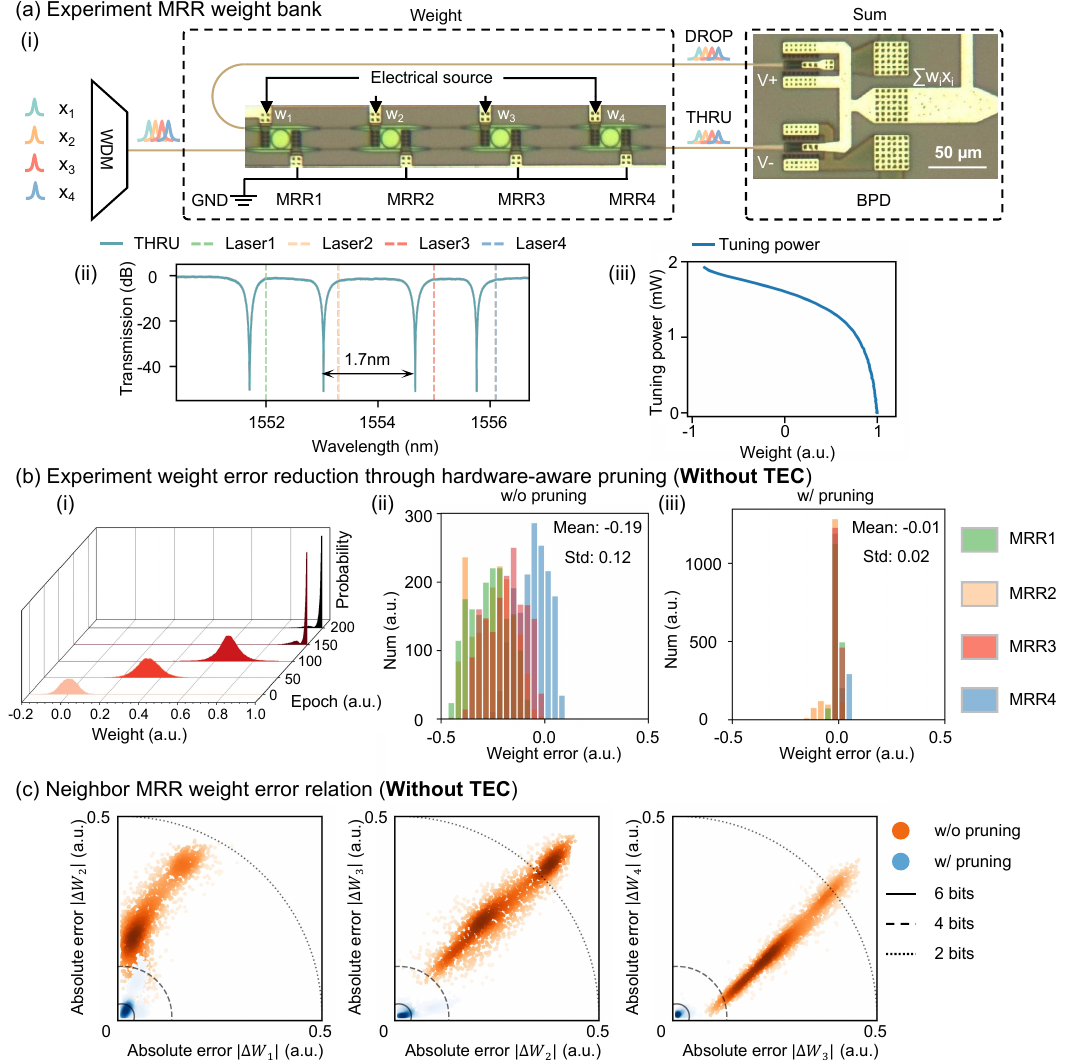}
\caption{(a-i) 4$\times$1 MRR weight bank perform dot product. Different modulated wavelength lights are combined by WDM, weighted by independent MRRs, and summed by BPD. (a-ii) MRR weight bank spectrum. (a-iii) Experiment measured tuning curve. (b-i) Weight distribution changes during optimization. (b-ii) Weight error distribution without the MRR pruning method. (b-iii) Weight error distribution with MRR pruning method. (c) Neighbor MRR weight error relation. WDM: Wavelength division multiplexer. MRR: Micro-ring resonator. GND: Ground. THRU: Through. BPD: Balanced photodetector. Std: Standard deviation. TEC: Thermoelectric controller.}
\label{fig:Weight_Error}
\end{figure*}

Our experiment first validates the effectiveness of our approach in reducing weight error on a two-layer CNN for MNIST digits classification. The first layer of the CNN comprises a convolution layer with 4 convolution kernels, each with a size of 3$\times$3. The output is activated by the ReLU nonlinear activation function, succeeded by a 2$\times$2 max-pooling operation. The subsequent layer is a fully connected layer with a size of 676$\times$10, mapping 10 categories of handwritten digits (0 to 9). We begin by training the CNN on a computer without incorporating a regularization term. This obtains 96.5\% training accuracy on the MNIST test dataset comprising 10,000 pictures. The trained weights are subsequently implemented on a MRR weight bank, as illustrated in Fig.\ref{fig:Weight_Error}(a-i). The spectrum of four MRRs is depicted in Fig.\ref{fig:Weight_Error}(a-ii). The resonant wavelengths of four MRRs are situated at 1551.7, 1553.0, 1554.7, and 1555.8 nm, respectively. The MRR weight bank design and fabrication are detailed in the supplement document. The implementation process goes by flattening the weight matrix and then adding tuning currents to MRRs sequentially based on their tuning curves.

During the implementation of weights, the output wavelengths of four lasers are initially adjusted to the off-resonance regions of four MRRs, at 1552.0, 1553.3, 1555.0, and 1556.1 nm, respectively. These lasers are modulated by signals generated by an arbitrary waveform generator (AWG) using 4 Mach-Zehnder modulators (MZMs). The modulated signals are combined by a wavelength division multiplexer (WDM) and then directed into the MRR weight bank through a grating coupler. Subsequently, the output signal, which is the photocurrent generated in BPD, goes into an off-chip bias-tee and transimpedance amplifier (TIA). Then, the signal is sampled and digitized by a real-time scope.

Before implementing the weights, we sweep the actuating current of each MRR to obtain a lookup table between the actuating current and weight. In the meanwhile, we record the power consumption for each weight, shown in Fig.\ref{fig:Weight_Error}(a-iii). Due to thermal drift, the weights deviate from the target weights. The actual weights implemented on the four MRRs are obtained through the pseudo-inverse method~\cite{zhang2022silicon}. We define weight error as:

\begin{equation}\label{eq: error}
  \Delta w = w_{true} - w_{target}
\end{equation}
where $w_{true}$ is the obtained weight value, $w_{target}$ is the target weight value.

TEC is a power-consuming module. In the experiment, we do not use TEC to show our approach can realize control-free operation. We implement 6760 trained weights in the two-layer CNN experimentally without TEC (Details are illustrated in the supplement document). The initial weight mainly distributes around 0, as is shown in Fig.\ref{fig:Weight_Error}(b-i). We measure the actual weight value using the pseudo-inverse method and calculate the error associated with each weight, generating a histogram represented in Fig.\ref{fig:Weight_Error}(b-ii). 

For weights without pruning, the mean value of weight error is -0.19, and the standard deviation of weight error is 0.12 for all weights, primarily due to a resonance drift at around 12 pm caused by thermal fluctuations (Details are illustrated in the supplement document). To mitigate the error, we re-train the CNN using the proposed pruning method, and implement the trained weights on the 4$\times$1 MRR again. Weight distribution changes clearly during training, shown in Fig.\ref{fig:Weight_Error}(b-i). Around 55.4\% weights are pruned to 1. As shown in Fig.\ref{fig:Weight_Error}(b-iii), after pruning, the weight error histogram converges around 0, the mean value increases to -0.01 and the standard deviation decreases to 0.02. The results are obtained without TEC under similar thermal drift conditions.

In our experimental setup, the primary cause of thermal drift stems from temperature fluctuations, because the duration of the experiment extends over several hours. As depicted in Fig.\ref{fig:Weight_Error}(c), the 2D-relation figures illustrate the absolute error between neighboring MRR weights. They show a significant 4-bit enhancement in weight error precision without TEC (Details are illustrated in the supplement document). A strong linear correlation is evident among weight errors without pruning. It affirms that external temperature fluctuations play a more substantial role than thermal crosstalk. When the MRR weight bank's size is larger and thermal crosstalk significantly affects the weight error, we anticipate our method could also reduce the thermal crosstalk due to most weights being achievable without actuating the current.

\begin{figure*}[ht]
\centering
\includegraphics[width=1\linewidth]{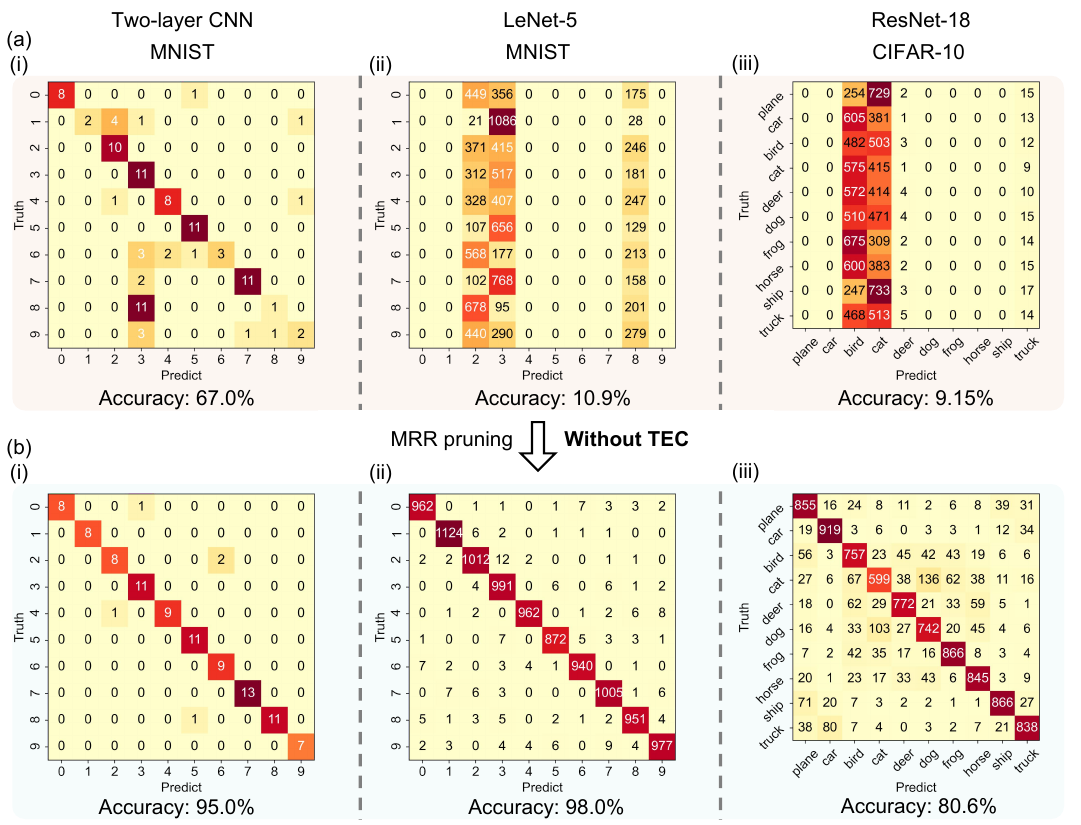}
\caption{Experiment inference confusion matrices under experiment error. (a-i) Two-layer CNN 'MNIST' classification confusion matrix without pruning. (a-ii) LeNet-5 'MNIST' classification confusion matrix without pruning. (a-iii) ResNet-18 'CIFAR-10' classification confusion matrix without pruning. (b-i) Two-layer CNN 'MNIST' classification confusion matrix with pruning. (b-ii) LeNet-5 'MNIST' classification confusion matrix with pruning. (b-iii) ResNet-18 'CIFAR-10' classification confusion matrix with pruning. MRR: Micro-ring resonator. TEC: Thermoelectric controller.}
\label{fig:accu}
\end{figure*}

Based on the prior experiment, we have confirmed the significant enhancement of our approach in strengthening MRR weight bank's resilience to resonance drift. Building upon this, we conduct another experiment to evaluate the improved inference accuracy of CNN. The inference is executed layer by layer. For the convolution layer, the 'im2col' method is applied to transform the convolution operation into matrix multiplication~\cite{chetlur2014cudnn}. This allows us to decompose large matrix multiplications into small dot products implementable on the MRR weight bank. Subsequently, the results from the first layer are modulated onto light as the input for the second layer. A similar matrix multiplication decomposition is performed for the second layer, which is implemented on the MRR weight bank. Throughout this experiment, no external TEC is used. The inference results, showcased in Fig.\ref{fig:accu}(a-i) and (b-i), reveal that the accuracy without our approach is merely 67.0\%. In contrast, after applying our approach, the experimental accuracy increases to 95.0\%, close to the theoretical value. This substantial improvement is attributed to the marked reduction in weight error and thermal crosstalk and an enhancement in the SNR.

\subsubsection{Extension to larger and deeper NNs and various machine learning tasks}
Moreover, we extend the application of our approach to larger and deeper NNs to validate its effectiveness with increased NN size. Initially, we utilize the LeNet-5 architecture with 60,000 weights for MNIST handwritten digits classification. The training dataset has 60,000 MNIST images, and the testing dataset has 10,000 images. Training the network without incorporating a regularization term yields a classification accuracy of 98.8\%. 

To add experimental error, we introduce weight error by randomly sampling from the histogram of experimental weight error. The error is reshaped to match the NN's size and added to the NN for dataset inference. The weight error leads to a drastic reduction in classification accuracy to 10.9\%. After applying our approach, around 92.1\% weights are pruned to 1. Weight error reduction contributes to an impressive improvement in the classification accuracy to 98.0\%. Subsequently, we extend our method to a larger and deeper NN, ResNet-18, which comprises approximately 12 million weights. We employ this NN for a more challenging image classification task, the CIFAR-10 dataset which comprises 60,000 32$\times$32 color images in 10 different classes. We select 50,000 images randomly for training and reserve 10,000 for testing. The trained accuracy without any experimental error is 82.7\%. While it decreases to 9.15\% even only introducing one-eighth of the experimental weight error. Our approach improves the classification accuracy to 80.6\%, comparable to the theoretical accuracy, in the presence of one-eighth of the experimental weight error. Our experiments show the general applicability of our approach across various NN sizes and machine learning tasks.

\begin{figure}[t]
\centering
\includegraphics[width=0.75\linewidth]{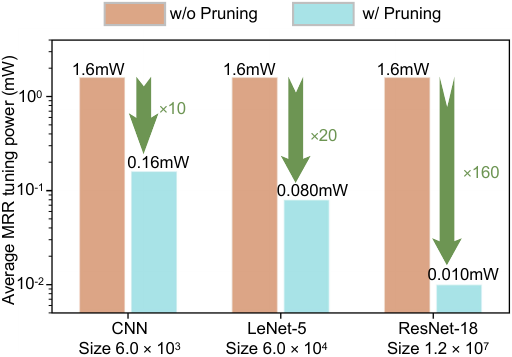}
\caption{Average MRR tuning power across different size neural networks. MRR: Micro-ring resonator. CNN: Convolutional neural network.}
\label{fig:power}
\end{figure}

\subsubsection{Power consumption reduction}
Finally, we evaluate the power consumption reduction achieved through our method. Larger NN allows more weights to be pruned to "1", thus leading to a more pronounced power consumption reduction. We denote the average MRR tuning power as the total tuning power over the number of weights. We find that differently sized NNs have similar average tuning power at 1.6mW. After applying our approach, the average tuning power is reduced by a factor of 10, 20, and 160 for three types of NNs, respectively, as shown in Fig.\ref{fig:power}.

\subsection{Application to other MRR-based PNNs}

\begin{table}[ht]
\centering
\caption{\bf Simulation results on Single-end and Crossbar MRR array}
\begin{tabular}{cccc}
    \hline
    & & w/o pruning & w/ pruning \\
    \hline
    \multirow{2}{*}{Single-end}&Accuracy &40.6$\%$&97.4$\%$ \\
     &Tuning power&2.1mW&0.14mW \\
    \hline
    \multirow{2}{*}{Crossbar}&Accuracy&74.2$\%$&97.4$\%$ \\
     &Tuning power&1.2mW&0.97mW \\
\hline
\end{tabular}
  \label{tab:Generality}
\end{table}

\begin{figure*}[ht]
\centering
\includegraphics[width=1\linewidth]{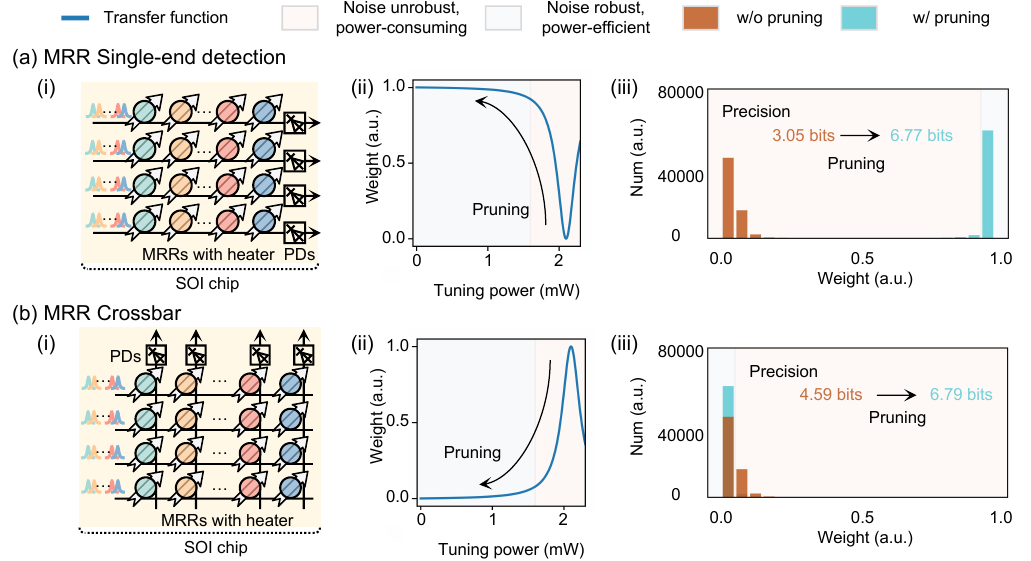}
\caption{Application to other integrated PNNs. (a-i) Diagram of Single-end detection MRR array. (a-ii) MRR Through port transmission tuning curve. (a-iii) Absolute weight distribution contrast, estimated wavelength drift at 50 pm. (b-i) Diagram of Crossbar MRR array. (b-ii) MRR Drop port transmission tuning curve. (b-iii) Absolute weight distribution contrast, estimated wavelength drift at 150 pm. MRR: Micro-ring resonator. PD: Photodetector. BPD: Balanced photodetector. SOI: Silicon on insulator. }
\label{fig:generality}
\end{figure*}

Previous experimental results have demonstrated that our approach increases the robustness of PNN and enhances precision in MRR-based PNN with BPD. To further validate our method's applicability across various MRR-based PNNs, we conduct simulations on two distinct architectures. One architecture is an MRR array with single-end detection, and the other is an MRR array with a crossbar structure, as depicted in Fig.\ref{fig:generality}(a-i) and (b-i). In these two architectures, negative weights and positive weights are separated to perform matrix multiplication independently~\cite{feldmann2019all,zhang2024compact,ohno2022si}. 

Fig.~\ref{fig:generality}(a-ii) shows the Through port transmission tuning curve in the single-end detection MRR array. The tunable weight values range from 0 to 1. Through conventional training methods, these weights tend to concentrate around 0, corresponding to the region that is sensitive to noise and consumes more power. Similar to the MRR weight banks with BPD, the region weight value of "1" remains resilient to resonance drift caused by thermal fluctuations. Therefore, we prune the weights to "1" using the revised loss function Eq.~\ref{eq: regularization}.

In the crossbar MRR array, the weights range is also from 0 to 1. However, the signal outputs from the Drop port, therefore, the noise-robust region changes from weight value "1" to weight value "0" as depicted in Fig.~\ref{fig:generality}(b-ii). In this case, we slightly modify Eq.~\ref{eq: regularization_slope} into Eq.~\ref{eq: regularization_mrr}
\begin{equation}\label{eq: regularization_mrr}
  L = f(y, \hat{y})+\alpha||\frac{\partial W}{\partial \lambda}||_2
\end{equation}
where $\frac{\partial W}{\partial \lambda}$ is the slope of MRR spectrum.

We employ the LeNet-5 architecture for classifying MNIST handwritten digits to evaluate the resilience of various MRR-based PNNs. In weight banks featuring single-end detection (Fig.~\ref{fig:generality}(a-iii)), we successfully prune the weights to '1'. Conversely, in MRR crossbar arrays, more weights are compressed to '0' compared to arrays without pruning, as illustrated in Fig.~\ref{fig:generality}(b-iii). In both cases, the computing precisions are improved with the additional benefit of power consumption.

To quantify the improvement, we emulate the resonance drift caused by thermal fluctuation and introduce sensitivity-related Gaussian noise $\Delta w_{noise} \sim N(0, |\frac{dw}{d\lambda}*\Delta \lambda|^2)$ to evaluate the robustness, where $\frac{dw}{d\lambda}$ represents the slope of the MRR spectrum, $\Delta \lambda$ is the estimated resonance drift. (Details are derived in the supplement document).

Our simulation results show significant control precision improvement both in the MRR weight bank with single-end detection and the crossbar MRR array, as shown in Fig.\ref{fig:generality}(a-iii) and (b-iii). In the MRR weight bank with single-end detection, the computing precision increases from 3.05 bits to 6.77 bits,  boosting the MNIST classification accuracy from 74.2\% to 97.4\%, the average MRR tuning power is reduced by 15 times, as depicted in Table.\ref{tab:Generality}. The result is estimated assuming a 50 pm resonance drift. In contrast, the MRR crossbar array exhibits higher control precision without pruning compared to the MRR weight bank with single-end detection because weights have been centered around 0. With pruning, weights can get closer to 0, resulting in an improvement in computing precision from 4.59 bits to 6.79 bits. The enhancement in computing precision leads to improvement in classification accuracy from 74.2\% to 97.4\%. This result is estimated at a 150 pm resonance drift level.

\section{PCM-based PNNs}
\label{section:PCM}

\begin{figure*}[ht]
\centering
\includegraphics[width=1\linewidth]{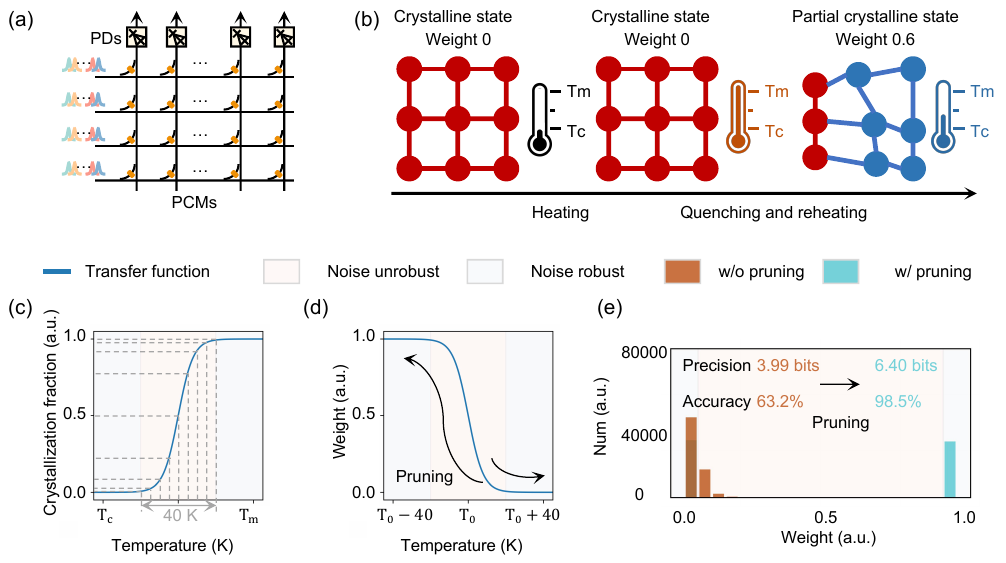}
\caption{PCM-based PNNs. (a) Diagram of PCM array. (b) Schematic diagram of PCM weight tuning. (c) Relationship between crystallization fraction and quenching temperature. The dashed lines show 8 level distinguishable weight levels with 5 K temperature fluctuation. (d) Relationship between weight and quenching temperature. (e) Absolute weight distribution contrast, estimated temperature fluctuation at 5 K. PCM: Phase change material. PD: Photodetector. }
\label{fig:pcm}
\end{figure*}

PCM-based PNNs achieve tunable weights by thermally adjusting the proportion of crystalline and amorphous states in the PCM~\cite{li2024performance}. In the crystalline state, most of the incoming light is absorbed, corresponding to a weight value of "0". In the amorphous state, most of the light is transmitted, representing a weight value of "1"~\cite{feldmann2021parallel}. Therefore, weight tuning is achieved by heating the PCM beyond melting temperature $T_m$ and then rapidly quenching, followed by heating the PCM to various temperatures between crystallization temperature $T_c$ and melting temperature $T_m$. In this way, PCM will be partially crystallized to a certain optical state; this corresponds to the specific weight value between 0 and 1, as shown in Fig.~\ref{fig:pcm}(b)~\cite{wei2023electrically}. However, imperfections, especially temperature fluctuations, will affect the crystallization fraction and introduce weight error. These imperfections also lead to limited resolvable weight levels and reproducibility caused by various factors, including material degradation, stochasticity in the crystallization process, and fluctuations in experimental switching~\cite{li2024performance}. Take $Sb_{2}Se_{3}$ as an example, Fig.\ref{fig:pcm}(c) shows crystallization fraction changes with quenching temperature for a heating duration of 1ms \cite{li2024performance}. The curve exhibits a sigmoid function, indicating the state changing from a pure crystalline state to an amorphous state. Our simulation results show that only 5 K thermal fluctuation, corresponding to 8 distinguishable levels, will reduce MNIST classification from 99.0\% to 63.2\% in the PCM crossbar array~\cite{feldmann2021parallel}.

To solve this problem with our approach, we first analyze the stable region of the PCM-based PNNs. We can observe that both the regions of pure crystalline state and amorphous state are insensitive to temperature fluctuation. The crystallization fraction can also be linearly transformed into weight values. The relation between weight and quenching temperature is shown in Fig.\ref{fig:pcm}(d). We use the data in reference \cite{li2024performance} to draw the specific numerical curve. We define $T_{0}$ as the intermediate temperature between $T_{c}$ and $T_{m}$, and the difference between $T_{c}$ and $T_{m}$ is about 80 K. 

To solve the inaccuracy problem of PCM, our approach is to relocate most weights to these two stable regions during training. We slightly modify Eq.\ref{eq: regularization_slope} into Eq.\ref{eq: regularization_pcm}, and optimize the NN with Eq.\ref{eq: regularization_pcm}.
\begin{equation}\label{eq: regularization_pcm}
  L = f(y, \hat{y})+\alpha||\frac{\partial W}{\partial T}||_2
\end{equation}
where $\frac{\partial W}{\partial T}$ is the slope of weight to the temperature change. This loss function forces most weights to relocate to the stable region where $\frac{\partial W}{\partial T} \to 0$. 

The weights are initialized with a uniform distribution of -1 to 1 to avoid only relocating weights to the weight 0 region. The weight distribution after pruning is shown in Fig.\ref{fig:pcm}(e). We can see most weights are successfully relocated to "0" and "1". 

To simulate the impact of thermal fluctuations and assess the robustness of our approach, we add noises in weight $\Delta w_{noise} \sim N(0, |\frac{dw}{dT}*\Delta T|^2)$ caused by temperature fluctuations. The simulation result is estimated assuming 5 K thermal fluctuation, 8 distinguishable weight levels. The simulation results (Fig.~\ref{fig:pcm}(e) shows that after pruning, most weights are relocated to noise-robust regions, corresponding to the crystalline state and the amorphous state of PCM. The weight precision improves from 3.99 bits to 6.40 bits. The MNIST classification accuracy improves from 63.2\% to 98.5\%.

\section{Conclusion}

In summary, we propose and demonstrate control-free and energy-efficient photonic computing through hardware-aware training and pruning. This approach is helpful for solving accuracy problems in PNNs. Our method can boost control precision by 4 bit, leading to a significant increase in PNN inference accuracy. Furthermore, our approach concurrently reduces power consumption associated with tuning device states to match desired weights, all achieved without needing cumbersome control algorithms or introducing additional hardware complexity. Our work presents an important step towards practical, energy-efficient, and noise-resilient implementation of large-scale PNNs. Furthermore, our novel idea of training the parameters of a physical neural network towards its noise-robust and energy-efficient region can be broadly applied to other physical neural networks, addressing the noise issue commonly faced by such systems.
We envision that future work can explore the integration of our training method with in-situ training ~\cite{wright2022deep, pai2023experimentally, perez2020multipurpose, huo2023optical,muller2023artificial,filipovich2022silicon} to further enhance the robustness and efficiency of photonic computing. The limit of our approach is the computation cost when the PNN system becomes complex. For example, in MZI-based PNNs each element within the linear transmission matrix is simultaneously determined by several variables rather than independently determined by each variable. The Mote-Carlo method and perturbation method can be applied to simulate such systems and find the stable regions~\cite{levy2020large,wu2020adversarial}.

\bibliography{sn-bibliography}

\section*{Acknowledgements}

This work was supported by ITF ITS/237/22, RGC YCRF C1002-22Y, RNE-p4-22 of the Shun Hing Institute of Advanced Engineering, NSFC/RGC Joint Research Scheme N\underline{\hspace{0.5em}}CUHK444/22, and CUHK Direct Grant 170257018, 4055143.

\section*{Author contributions}

T.X. and C.H. conceived the ideas. T.X. performed simulations. T.X., W.Z., J.Z. designed the experiment and conducted the experimental measurements. T.X., Z.L., Q.X., B.W. and M.L. analyzed the results. T.X., W.Z., J.Z., B.J.S. and C.H. wrote the manuscript. X.X., B.J.S. and P.R.P. provided theoretical and experimental advice and revised the manuscript. C.H. supervised the research and contributed to the general concept and interpretation of the results. All
the authors discussed the data and contributed to the manuscript.

\end{document}